Correspondence to:
G. Torri,
torri@fas.harvard.edu

# Mechanisms for convection triggering by cold pools

Giuseppe Torri[1], Zhiming Kuang[1,2], and Yang Tian[1]

[1]Department of Earth and Planetary Sciences, Harvard University, Cambridge, Massachusetts, USA, [2]School of Engineering and Applied Sciences, Harvard University, Cambridge, Massachusetts, USA

**Abstract** Cold pools are fundamental ingredients of deep convection. They contribute to organizing the subcloud layer and are considered key elements in triggering convective cells. It was long known that this could happen *mechanically*, through lifting by the cold pools' fronts. More recently, it has been suggested that convection could also be triggered *thermodynamically*, by accumulation of moisture around the edges of cold pools. A method based on Lagrangian tracking is here proposed to disentangle the signatures of both forcings and quantify their importance in a given environment. Results from a simulation of radiative-convective equilibrium over the ocean show that parcels reach their level of free convection through a combination of both forcings, each being dominant at different stages of the ascent. Mechanical forcing is an important player in lifting parcels from the surface, whereas thermodynamic forcing reduces the inhibition encountered by parcels before they reach their level of free convection.

## 1. Introduction

In recent years, different authors have suggested that the subcloud layer is a crucial component in the rapid transition from shallow to deep convection [*Tompkins*, 2001; *Chaboureau et al.*, 2004; *Khairoutdinov and Randall*, 2006; *Kuang and Bretherton*, 2006; *Zhang and Klein*, 2010; *Böing et al.*, 2012]. In particular, *Böing et al.* [2012] focused on the positive feedback between the organization of the subcloud layer and the deepening of clouds and showed that modifying certain properties of the former can inhibit convection deepening.

The organization of the subcloud layer proceeds through the formation of cold pools, areas of evaporatively cooled downdraft air that spread on the surface as density currents underneath precipitating clouds [*Droegemeier and Wilhelmson*, 1985a]. As noted by various authors [e.g., *Tompkins*, 2001; *Schlemmer and Hohenegger*, 2014; *Li et al.*, 2014], the edges of cold pools contain air parcels in which falling precipitation evaporated in the subcloud layer and, accordingly, are colder and moister than the horizontal average. This can be appreciated in Figure 1, which shows the horizontal distribution of potential temperature and specific humidity near the model surface during the simulation discussed in this letter (see section 2 for more details). The innermost regions of cold pools are colder and drier than the average and are represented by blue colors in both panels. The edges of cold pools are characterized by a negative temperature anomaly, but the dark red colors that represent them in Figure 1 (right) imply that they are richer in water vapor than the average.

It has been long known that cold pools can trigger new convective cells. The forced lifting of air along the gust front of a cold pool was one of the first mechanisms to be identified as responsible for convection triggering [*Purdom*, 1976; *Weaver and Nelson*, 1982]. Such lifting could be the result, for example, of the interaction of a cold pool with low-level wind shear, as is typically the case in organized systems like multicell storms or squall lines [see, for example, *Knupp and Cotton*, 1982; *Rotunno and Klemp*, 1985; *Rotunno et al.*, 1988; *Weisman et al.*, 1988; *Weisman and Rotunno*, 2004] or could be caused by the collision of two cold pools [e.g., *Droegemeier and Wilhelmson*, 1985a, 1985b]. This mechanism will be referred to as *mechanical forcing*.

More recently, it has been suggested that cold pools can also contribute to the generation of new deep convective clouds through another mechanism [*Betts*, 1984; *Tompkins*, 2001]. Considering a series of simulations, *Tompkins* [2001] observed that cold pools could develop also in environments with low vertical wind shear, a condition often found over the tropical ocean. As also noted by other authors [*Young et al.*, 1995; *Saxen and Rutledge*, 1998], an important difference with cases with strong wind shear is that new clouds seem to form when cold pools from which the air parcels originated are in a mature or even decaying stage. At this point, the gust front is not strong enough to lift air mechanically, so that mechanical forcing is unlikely to play a role in this case. Instead, Tompkins suggested that the accumulation of water vapor found in the boundaries of spreading cold pools may provide enough buoyancy to air parcels that even a small lifting could be sufficient to initiate convection. To highlight the importance of the thermodynamic state of air parcels forced in this way, this mechanism will be called *thermodynamic forcing*.

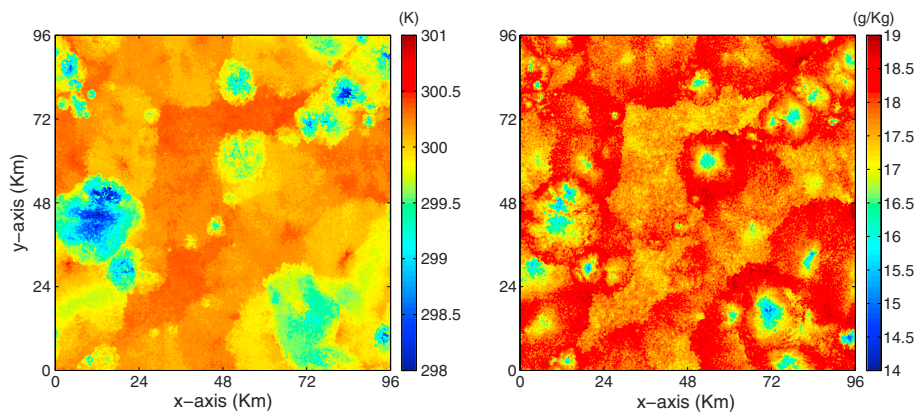

**Figure 1.** Horizontal sections of (left) potential temperature and (right) water vapor specific humidity at 25 m from the model surface.

Although it is fair to expect both forcings to be relevant for spawning new convective elements, the relative importance of each in different scenarios is not known. In this letter, the problem is addressed using a method based on a Lagrangian Particle Dispersion Model and results from a simulation of radiative-convective equilibrium (RCE) over the ocean. In future work, the current analysis will be extended to a wider spectrum of scenarios.

## 2. Methods
### 2.1. Model and Simulations

The model used in this work is the System for Atmospheric Modeling (SAM), version 6.8.2, which solves the anelastic equations of motion and uses liquid water static energy, total nonprecipitating and precipitating water as thermodynamic prognostic variables [*Khairoutdinov and Randall*, 2003]. The equations are solved with doubly periodic boundary conditions in the horizontal directions. A prognostic turbulent kinetic energy (TKE) 1.5-order closure scheme is used to parameterize subgrid-scale effects, and a bulk microphysics scheme is chosen for simplicity. The surface fluxes are computed using the Monin-Obukhov similarity theory. The simulations are carried out over a domain of 96 × 96 $km^2$ in the horizontal directions and 41 km in the vertical directions. The grid size in the horizontal directions is 250 m, whereas it varies in the vertical directions: 50 m in the lowest 1 km, 125 m between 1 km and 5 km, 250 m between 5 km and 10 km, and 500 m for the other levels. The time step is 3 s.

The model is run with the Lagrangian Particle Dispersion Model (LPDM) discussed in *Nie and Kuang* [2012]. For every column in the domain, the LPDM is initialized with 480 particles, their positions being distributed randomly over the bottom 4 km of the model domain. Every 20 time steps, the positions of the Lagrangian particles and the three-dimensional model fields from SAM are recorded. A cross comparison of Lagrangian and Eulerian data provides the complete histories of thermodynamic and dynamical properties of the particles.

After setting the sea surface temperature at 302.65 K, the zenith angle at 50.5° with halved solar constant and the mean winds to zero, the model is run for 30 days without Lagrangian particles on a smaller horizontal domain measuring 64 × 64 $km^2$, until an RCE state is reached. The soundings at the end of this period are then taken to initialize the simulation with the Lagrangian particles that will be used in the following: after an initial spin up of 36 h, data are collected for 24 h of model time. No nudging is imposed on the horizontal winds. The mean precipitation rate during the last simulated day is 3.2 mm/d.

### 2.2. Tracking Algorithm

Since cold pools are essentially density currents, the ideal candidate for their identification is the density potential temperature [*Emanuel*, 1994], defined as

$$\theta_\rho = \theta \left[ 1 + \left( \frac{R_v}{R_d} - 1 \right) q_v - q_l \right], \qquad (1)$$

where $q_v$ and $q_l$ are, respectively, the specific humidities of water vapor and liquid water; $R_d$ is the gas constant for dry air, having value 287.04 J kg$^{-1}$ K$^{-1}$, and $R_v$ is that for water vapor and equals 461.5 J kg$^{-1}$ K$^{-1}$. The density potential temperature represents the potential temperature that dry air would have to possess the same density as the moist air in question.

In order to track the cold pools in the domain, a novel algorithm is introduced, which uses a combination of Eulerian and Lagrangian data. The algorithm searches for cold pools by scanning through the grid boxes of the subcloud layer, here defined as the collection of the bottom layers of the model for which the time average of the specific humidity of nonprecipitating cloud condensate, $q_n$, during the 24 h of simulation is below $10^{-5}$ kg/kg.

As a first step, when $\theta_\rho$ of a particle is lower than the horizontal average by 1 K, it is considered a *cold pool particle*: the algorithm starts tracking it and assigns it a clock initialized at zero. The threshold of 1 K has been chosen by eye inspection of the model outputs, and sensitivity tests have been carried out to verify that results presented here are not particularly sensitive to it. The tracking continues and the clock is increased until the particle is neutrally buoyant.

If more than 50% of particles in a grid box are cold pool particles, the grid box is considered a *cold pool grid box*. In the three-dimensional domain, every ensemble of six connected cold pool grid boxes is identified as a *cold pool*. If a particle in the subcloud layer enters a cold pool grid box, then the particle is considered an entrained cold pool particle and it is given a clock with a value equal to the mode of the other particles' clocks in the grid box. The mode is interpreted as the age of a cold pool grid box, and the collection of modes of the grid boxes of a cold pool can provide an estimate of the *age of the cold pool* itself. When a cold pool particle recovers its buoyancy, the value of its clock provides a measure of the total amount of time spent inside the cold pool, hereinafter *residence time*.

The algorithm also considers particles in the surrounding of cold pools, which are likely to be lifted mechanically by the cold pool's gust front. When a particle is within a distance of 1 km from a cold pool grid box, it is considered by the algorithm as being in *proximity* to a cold pool. Particles that are neither within nor in proximity to a cold pool will be referred to as *environment particles* for the reminder of this letter.

To each particle in proximity of a cold pool, the algorithm associates a clock by considering all the cold pool grid boxes within a distance of 1 km from the particle: the mode of the distribution of clocks found in this area is assigned to the particle's clock. This way, the algorithm can provide an estimate of the age of the cold pool that lifted the particle. The value of 1 km for the proximity seems reasonable for the case at hand and coincides with the value used in *Li et al.* [2014] to define the cold pool ambient region, a similar concept to the proximity introduced above.

### 2.3. Forcing Signatures

In the present study, the main sources of information to distinguish between the forcings outlined in section 1 are given by the particles' histories. Whenever a particle reaches its level of free convection (LFC), defined by the condition that the particle is in a grid box with positive buoyancy and $q_n$ higher than $10^{-5}$ kg/kg, its positions at previous time steps are considered from the moment the particle acquired positive vertical velocity. Out of all the particles that reached their LFC, only those that started ascending in the subcloud layer are considered; this will prevent particles entrained within clouds to interfere with the analysis. The selected particles will be referred to as *triggered Lagrangian particles*.

Combining the positions of each particle during its ascent with the Eulerian outputs obtained from the cloud-resolving model (CRM), it is possible to reconstruct the histories of accelerations of the particle; the current study focuses on pressure gradients, mechanical and buoyancy-driven, $p_M$ and $p_B$, and buoyancy, $B$. The first two are defined as the solutions of the diagnostic equations [*Houze*, 1994]:

$$\nabla^2 p_M = -\nabla \cdot \left( \rho \vec{v} \cdot \nabla \vec{v} \right) \text{ and} \qquad (2)$$

$$\nabla^2 p_B = \partial_z (\rho B), \qquad (3)$$

where $\vec{v}$ is the velocity vector. Buoyancy is defined as

$$B = g\left(\frac{\theta_\rho - \overline{\theta_\rho}}{\overline{\theta_\rho}}\right), \tag{4}$$

where $g$ is the gravitational acceleration and $\overline{\theta_\rho}$ is the horizontal average of density potential temperature.

Buoyancy can be considered as a good indicator of the thermodynamic effect of cold pools: particles coming from the moisture anomaly at the leading edge of older cold pools are likely to experience large positive values of buoyancy as they start ascending. However, in order to assess whether the thermodynamic advantage provided by cold pools plays any role in the subcloud layer, buoyancy pressure gradients have to be considered as well [e.g., *Davies-Jones*, 2003]. Buoyancy-driven pressure perturbations can be thought of as an indication of the work that has to be done by a parcel to move surrounding air out of the way during its ascent and they typically counterbalance buoyancy accelerations [*Houze*, 1994]. The sum of buoyancy and buoyancy pressure gradients, hereafter *total buoyancy*, will indicate the effective importance of the thermodynamic forcing in triggering parcels.

In order to better illustrate the contribution of cold pools' fronts to the mechanical forcing of particles, the vertical components of mechanical pressure gradients of the triggered Lagrangian particles are compared with those of the environment grid boxes.

For the thermodynamic forcing, the triggered Lagrangian particles are compared with an ensemble of hypothetical parcels lifted from the environment to their LFC. First, the mean profiles of moist static energy (MSE) and total water specific humidity, $q_t$, of the triggered Lagrangian particles are used to determine the fractional entrainment rates at all heights using a bulk plume model. Then, the same plume model is initialized with values taken from the joint distribution of MSE and $q_t$ of the environment at the surface. Interpreting each value in the distribution as a hypothetical parcel, the bulk plume is used to reconstruct its thermodynamic state from the surface to its LFC. Comparing the average buoyancy profile obtained from the hypothetical parcels with that obtained from the triggered Lagrangian particles will provide an estimate of the thermodynamic forcing of cold pools at the surface and in the inhibition layer.

Next, the ages of cold pools are examined. Since the positive pressure anomaly at the leading edge of the cold pool becomes weaker as the density difference between cold pool and the environment diminishes, the gust fronts of older cold pools provide a weaker lifting than younger ones. At the same time, parcels in the leading edge of relatively old cold pools are moister than the subcloud layer average, are also less negatively buoyant than parcels in younger cold pools, and are thus thermodynamically favored for ascending. These two premises support the idea that the cold pool age when a particle is triggered can serve as an indicator of the nature of the forcing. Again, whenever a particle reaches its LFC, its history in the subcloud layer is examined: if the particle is within or in proximity to a cold pool during its ascent in the subcloud layer, then it is assigned the age of the cold pool as described above.

Finally, another variable that will be used in this analysis is the particles' residence time. Considering an idealized situation in which mechanical forcing is dominant, a substantial amount of particles lifted by the cold pools' fronts will likely come from the environment and their residence time in cold pools should thus be expected to be nil. On the other hand, parcels triggered thermodynamically are more likely to be coming from the cold pools themselves, precisely their leading edges, and should have higher residence times than in the mechanical forcing case.

## 3. Results

Discussion of the results begins by considering the accelerations of the particles from the time they acquired positive vertical velocity in the subcloud layer until they reach their LFC. Almost all particles in this category are either coming from a cold pool or they have been in proximity to one. The four panels of Figure 2 show the average accelerations as a function of height: the panels represent (top left) the mechanical pressure gradient and (top right) the total buoyancy, which is further divided in (bottom left) buoyancy pressure gradient and (bottom right) buoyancy.

There are two main features that stand out from the figure. The first one is that, compared to the total buoyancy of the parcels, the mechanical pressure gradient accelerations are bigger in the lowest part of the

...

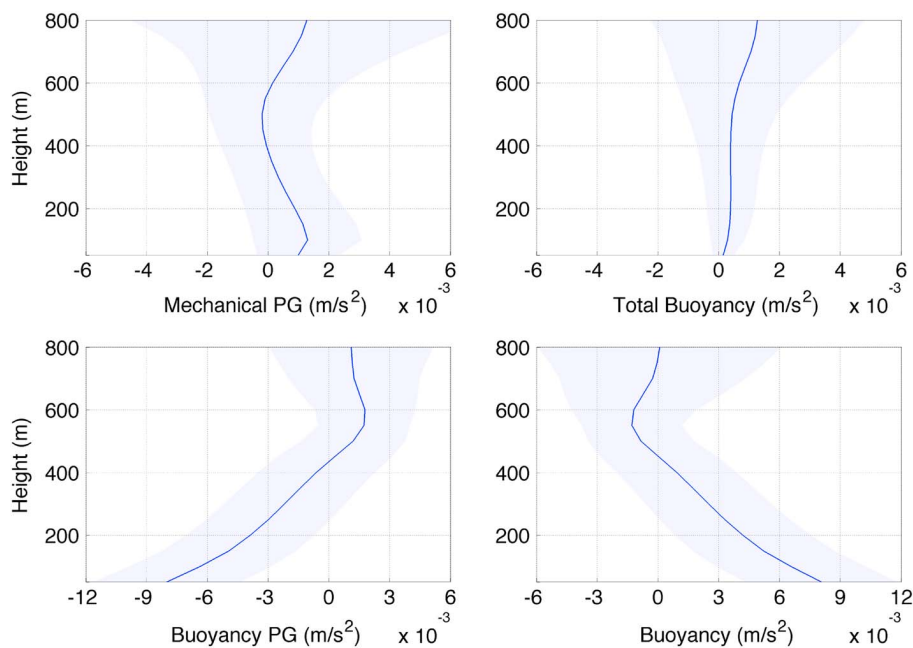

**Figure 2.** Profiles of average accelerations of the triggered Lagrangian particles. (top left) Mechanical pressure gradient and (top right) total buoyancy, and (bottom left) buoyancy pressure gradient and (bottom right) buoyancy. Shaded blue areas show values within a standard deviation from average.

subcloud layer at least by a factor 2 and are, therefore, the main contributing term in the particles' dynamical budget near the surface. Things change progressively as altitude increases and mechanical pressure gradients decrease steadily, reaching even small negative values in the convective inhibition layer. The relatively large values at the surface and the decay with height suggest that the mechanical pressure gradients of the particles are mainly due to the lifting by cold pools' fronts. To support this, Figure 3 shows a comparison between the distribution of mechanical pressure gradients of the triggered Lagrangian particles with those of the grid boxes of the environment. The distributions are built considering only the bottom 300 m of the domain, where Figure 2 suggests that cold pools' fronts would be most intense. The shapes of the curves are similar, although the one relative to the triggered Lagrangian particles is broader and more positively skewed. Most importantly, the distribution for the environment is centered around $2.1 \times 10^{-4}$ m/s$^2$, whereas that for the triggered Lagrangian particles is centered around $1.06 \times 10^{-3}$ m/s$^2$. This indicates that the large values of mechanical pressure gradients of the triggered Lagrangian particles are mostly due to the cold pool lifting.

The second interesting feature in Figure 2 is the degree of compensation between buoyancy and buoyancy pressure gradients in the subcloud layer, particularly near the surface, where buoyancy attains relatively large values. When buoyancy pressure gradients are taken into account, the increased buoyancy given by the positive moisture anomaly surrounding cold pools is almost entirely canceled out. From the dynamical point of view, total buoyancy begins to contribute significantly only in the upper part of the subcloud layer and proves to be a key factor in the convective inhibition layer. Part of the causes for the large values of buoyancy at the surface could be attributed to the fact that buoyancy is measured with respect to a reference state that includes also the negatively buoyant cores of cold pools. However, on average, the buoyancy of the triggered Lagrangian particles at the surface is still greater than that of the environment by $2.2 \times 10^{-3}$ m/s$^2$.

The thermodynamic forcing of cold pools is not limited to accelerations in the subcloud layer. It also extends above cloud base through its effect on the lifting condensation level and the convective inhibition encountered by ascending parcels. In order to better appreciate this, Figure 3 shows a comparison between the average buoyancy of triggered Lagrangian particles (blue) and the hypothetical average buoyancy of a parcel lifted from the environment (green). Although obtained from an idealized computation, the comparison indicates that the triggered Lagrangian particles experience a much smaller convective inhibition than environment parcels: −0.20 J/kg versus −1.02 J/kg. Notice that the total contributions of mechanical pressure gradients and total buoyancy to the kinetic energy at the bottom of the inhibition layer for the average triggered Lagrangian particle amount to 0.30 J/kg to and 0.18 J/kg, respectively. Thus, the reduction of convective inhibition by the thermodynamic effect of cold pools seems very important to allow a substantial number of particles to reach their LFC.

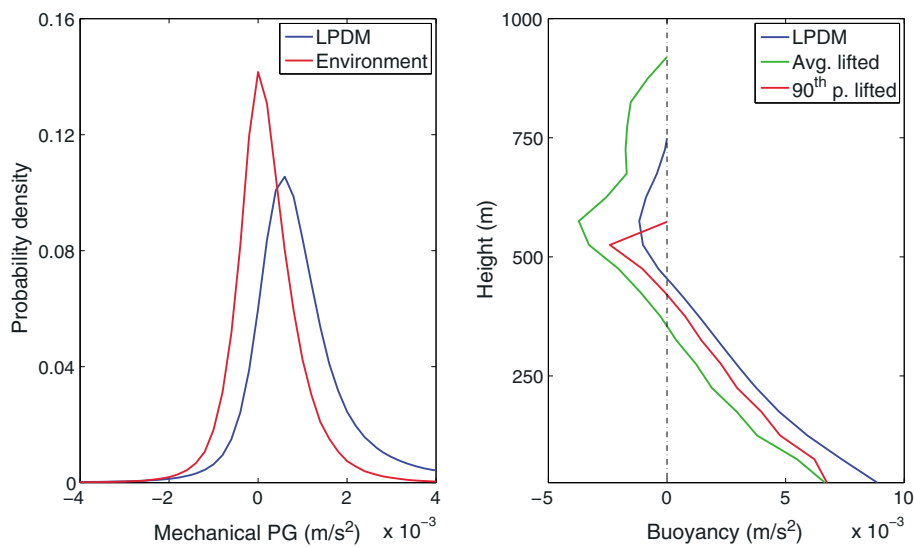

**Figure 3.** (left) A comparison between the distribution function of mechanical pressure gradients of triggered Lagrangian particles and of environment grid boxes in the bottom 300 m of the domain; (right) a comparison between the average buoyancy profile of triggered Lagrangian particles (blue) and two hypothetical parcels lifted from the surface. The parcels are initialized with the average environmental conditions (green) and with conditions corresponding to the 90th percentile of MSE (red).

The average moist static energy of triggered Lagrangian particles is greater than the environment by 0.76 K, and this difference is mostly contributed by water vapor: the former is colder by 0.01 K but moister by 0.31 g/kg. Thus, it seems that the moisture difference is the key element to reduce the convective inhibition encountered by the particles. In order to find environmental parcels that experience a similar inhibition to the triggered Lagrangian particles influenced by cold pools, it is necessary to consider very high percentiles of the distribution of properties. For example, Figure 3 also shows the buoyancy profile of parcels in the 90th percentile of the MSE distribution (red) at the surface, corresponding to a value of 346.54 K, greater than the average of triggered Lagrangian particles by 0.38 K. These parcels start off at the surface with almost the same buoyancy as the average parcels lifted from the environment but encounter an inhibition of −0.18 J/kg, which is comparable to that of the triggered Lagrangian particles. At the surface, the parcels in the 90th percentile of MSE are colder by 0.09 K than the average triggered Lagrangian particles and moister by 0.19 g/kg.

Next, the cumulative distribution of ages of cold pools that triggered particles is shown in Figure 4 (left). The curve grows rather quickly: 65% of the triggered Lagrangian particles have been lifted by cold pools that are younger than 4 h. This suggests that cold pools are relatively young when they lift particles and, therefore, supports the idea that mechanical forcing is an important player in the case examined.

Finally, the cumulative distribution of residence times shown in Figure 4 (right) indicates that, of all the triggered particles, 51% have never resided within a cold pool and 44% have resided for less than 4 h. The relatively large number of particles that have never been inside a cold pool strengthens the hypothesis of mechanical forcing being important in the case examined in this study.

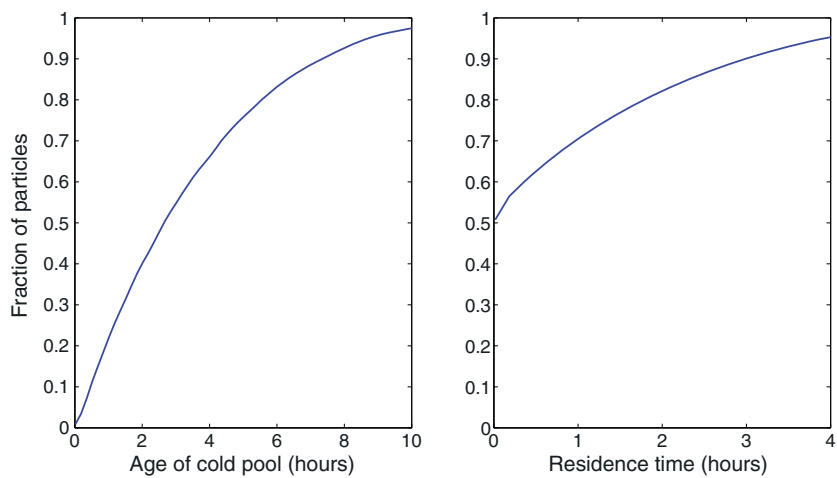

**Figure 4.** Cumulative distributions (left) of cold pool ages when triggered Lagrangian particles are lifted and (right) of the residence time of such particles.

## 4. Conclusions

This letter has focused on a better understanding of the importance of cold pools in triggering convection. In particular, care has been given to distinguish between the mechanical forcing, mainly due to gust front lifting, and the thermodynamic forcing, given by a positive anomaly of moist static energy surrounding cold pools. Results have been obtained with an LPDM, run using soundings from an RCE case over the ocean and with no wind shear. According to the arguments put forward by *Tompkins* [2001], the thermodynamic forcing should be the dominant triggering mechanism in this scenario.

In section 2, a distinction between forcings was proposed by looking at three signatures: the accelerations of triggered particles before they reach LFC; the age of the cold pools in the proximity of the ascending triggered particles and the time spent by said particles within cold pools.

It has been shown that triggered particles in the subcloud layer exhibit positive values of mechanical pressure gradients, especially close to the surface. A comparison with areas without cold pools has been used to support the hypothesis that these values are due to lifting by cold pools' gust fronts.

The results have also shown that, although the thermodynamic forcing is manifest in the large values of buoyancy of particles at the surface, the buoyancy pressure gradients cancel most of this advantage. The total buoyancy of the particles is much smaller than the mechanical pressure gradients at the surface but becomes gradually important close to the inhibition layer encountered by the particles. A comparison with a hypothetical set of parcels lifted from the environment has also suggested that the thermodynamic forcing of cold pools is still of great importance in reducing the inhibition encountered by particles that interact with cold pools.

Finally, it has been shown that cold pools tend to be relatively young when they lift triggered particles and that a large number of triggered particles reach their LFC without residing in a cold pool.

Therefore, the picture that emerges from the collection of results presented here is not one in which one forcing simply dominates over the other. Instead, parcels are lifted to the LFC through the cooperation of both forcings, each being important at different stages of the ascent. At the surface, since the buoyancy pressure gradients cancel to a large degree the high values of buoyancy of the parcels, mechanical lifting is needed for parcels to start ascending. With the kinetic energy provided by the cold pools' fronts and the total buoyancy, the parcels ascend through the subcloud layer and reach the convective inhibition layer. In this region, the effects of the thermodynamic forcing become manifest: because the parcels left the surface from areas of high moisture content, they experience a much smaller inhibition than would a parcel from the environment and can, therefore, reach their LFC more easily.

**Acknowledgments**
The authors thank Paul Edmon for precious and tireless assistance with the Harvard Odyssey cluster on which the simulations were run. The authors are grateful to three anonymous reviewers whose insightful comments helped improve the quality of the manuscript. This research was supported by the Office of Biological and Environmental Research of the U.S. Department of Energy under grant DE-SC0008679 as part of the ASR program. The data for this paper are available upon request.

The Editor thanks three anonymous reviewers for their assistance in evaluating this paper.